\documentclass[aps, prl, superscriptaddress,twocolumn,showpacs]{revtex4-2}
\usepackage{graphicx}
\usepackage{amsmath, amsthm, amssymb}
\usepackage{upgreek}
\usepackage{dcolumn}
\usepackage{bm}
\usepackage{hyperref}
\hypersetup{hypertex=true,
            colorlinks=true,
            linkcolor=red,
            urlcolor=black,
            citecolor=blue}

\begin{document}


\title{Size and Shape Fluctuations of Ultrasoft Colloids}

\author{Huarui Wu}
\affiliation{Department of Engineering Physics and Key Laboratory of Particle and Radiation Imaging (Tsinghua University) of Ministry of Education, Tsinghua University, Beijing 100084, China.}
 
\author{Jing Song}%
\affiliation{Department of Engineering Physics and Key Laboratory of Particle and Radiation Imaging (Tsinghua University) of Ministry of Education, Tsinghua University, Beijing 100084, China.}


\author{Wei-Ren Chen} 
 \affiliation{Neutron Scattering Division, Oak Ridge National Laboratory, Oak Ridge, Tennessee 37831, United States}

\author{Kun Song}%
\affiliation{Department of Engineering Physics and Key Laboratory of Particle and Radiation Imaging (Tsinghua University) of Ministry of Education, Tsinghua University, Beijing 100084, China.}

\author{Lionel Porcar}
 \affiliation{Institut Laue-Langevin, B.P. 156, F-38042 Grenoble CEDEX 9, France}
 
\author{Zhe Wang}
\email[Corresponding author: ]{zwang2017@mail.tsinghua.edu.cn}
\affiliation{Department of Engineering Physics and Key Laboratory of Particle and Radiation Imaging (Tsinghua University) of Ministry of Education, Tsinghua University, Beijing 100084, China.}


\begin{abstract}
Ultrasoft colloidal particle fluctuates due to its flexibility. Such fluctuation is essential for colloidal structure and dynamics, but is challenging to quantify experimentally. We use dendrimers as a model system to study the fluctuation of ultrasoft colloids. By considering the dynamic polydispersity in the small-angle neutron scattering (SANS) model, and introducing the fluctuation of invasive water into the contrast in SANS, we reveal the fluctuating amplitudes of the size and shape of the dendrimer of generation 6 at finite concentrations. The size fluctuation is suppressed while the shape fluctuation increases as the weight fraction of dendrimers passes 11\%. With neutron spin echo data, we suggest that such crossover originates from the competition between the inter- and intra-particle dynamics. Further investigation on lower-generation samples shows a contrary result, which suggests a structural basis for these dynamic phenomena.
\end{abstract}

\maketitle


Ultrasoft colloids, such as dendrimers and star polymers, are featured by extraordinary molecular flexibility \cite{Likos2006, Grest2007, Burchard1999, Winkler2014, Ballauff2004, Gast1996}. The elastic energy stored by such a particle that undergoes a large strain can be just hundreds of or even tens of the thermal energy \cite{Likos1998, Likos2001a, Likos2002, Vlassopoulos2014}, which distinguishes them from emulsions, most microgels, or other common deformable particles \cite{Princen1983, Weitz1995, Weitz2011, Bonnecaze2010, Vlassopoulos2014, Nieves2015, Richter1987, Narayanan2017}. Consequently, ultrasoft colloids exhibit significant molecular fluctuations. These fluctuations are crucial in many physical processes. Simulations suggest that the size and shape fluctuations considerably affect the self-diffusion of particles \cite{Maiti2009}. At concentrations close to the random close packing, it is proven that the shape fluctuation is related to the stress-releasing and -building of particle, and plays an essential role in the unusual dynamics \cite{Zaccarelli2019}. On the practical level, fluctuations can modulate the particle conformation \cite{Likos2003} that profoundly impacts a large variety of applications \cite{Yang2020, Zeng2011, Frechet1999, Kesharwani2014}. For instance, dendrimers have been established as drug carriers. The conformational fluctuation directly affects the size, shape and internal cavity of the molecule, which are important for the drug loading and release ability and the permeability across the biomembrane \cite{Jansen1995, Kitchens2005, Farideh2017, Tian2013, Goddard2009}. Therefore, there is a strong need to quantify the size and shape fluctuations to advance our understanding on both the physics and applications of ultrasoft colloids. 

In the past two decades, neutron spin echo (NSE) technique has been used to measure the intra-particle motions of ultrasoft colloids at the dilute limit, and the fluctuations of the structural unit and shape of the isolated particle were revealed \cite{Farago2003, Rathgeber2006}. While many ultrasoft colloidal suspensions of physical and technological importance are with finite concentrations \cite{Vlassopoulos2014, Zaccarelli2019, Foffi2013}. For these systems, the NSE analysis on the molecular fluctuation is much complicated by the difficulty in experimentally determining the collective translational diffusion of particles and its coupling to the intra-particle dynamics \cite{Nagele2010}. On the other hand, the fluctuations influence the distribution of the particle conformation, which could be reflected in small-angle scattering (SAS) patterns. Note that, most previous SAS studies of ultrasoft colloids have not explicitly considered the effect of particle size and shape fluctuations \cite{Likos2006, Liu2012}. Many SAS analyses deteriorate at volume fractions higher than about 10\%. For example, the calculated SAS curve may not well match the position or height of the main peak \cite{Likos1998, Likos2002}, or underestimate the intensity at small $Q$ ($\bm{Q}$ is the scattering vector) \cite{Gupta2015a}. Thus, it is possible that the fluctuation effect is important in interpreting the SAS data of ultrasoft colloids, and can be extracted by reasonable modelling. 

In this letter, we investigate the size and shape fluctuations of ultrasoft colloids by using neutral poly(amido amine) (PAMAM) dendrimers  dissolved in water as the model system. Six concentrations, $c=1$, 5, 10, 12.5, 15, and 20 wt\% (weight fraction of dendrimers) were measured. We first introduce an approach to examine these fluctuations in the dendrimer molecule by small-angle neutron scattering (SANS). It will be seen that explicitly considering the size fluctuation is of special importance in SANS analysis. Then, by combining the NSE result, and investigating lower-generation samples, we reveal the mechanism for the dependences of these fluctuations on the molecular structure and dynamics.

Dendrimers with the same molecular weight possess different conformations due to molecular fluctuations. Thus, dendrimer solutions can be regarded as polydisperse. We introduce the \textit{dynamic polydispersity} \cite{Rathgeber2002} to reflect this fluctuation effect, which leads to the following expression of the SANS intensity \cite{Chen1983}:
\begin{equation}
I(Q)=n_{\textrm{p}} A P(Q) S'(Q),\label{eq:IQ_1}
\end{equation}
where $n_{\textrm{p}}$ is the number density of dendrimer molecules, $A$ denotes the contrast of the scattering length between solute particle and solvent, $P(Q)$ is the average form factor normalized at $Q=0$, and $S'(Q)$ is the apparent structure factor given by $S'(Q)=1+\beta(Q)[S(Q)-1]$, where $\beta(Q)$ is the polydispersity factor \cite{Chen1983} that incorporates the size and shape fluctuations into the analysis, and $S(Q)$ is the inter-particle structure factor. $S(Q)$ is calculated by the Percus-Yevick closure of the Ornstein-Zernike equation \cite{Hansen2013} with a Gaussian pair potential \cite{Likos2002}.

\begin{figure}[h]
	\centering
	\includegraphics[scale=1]{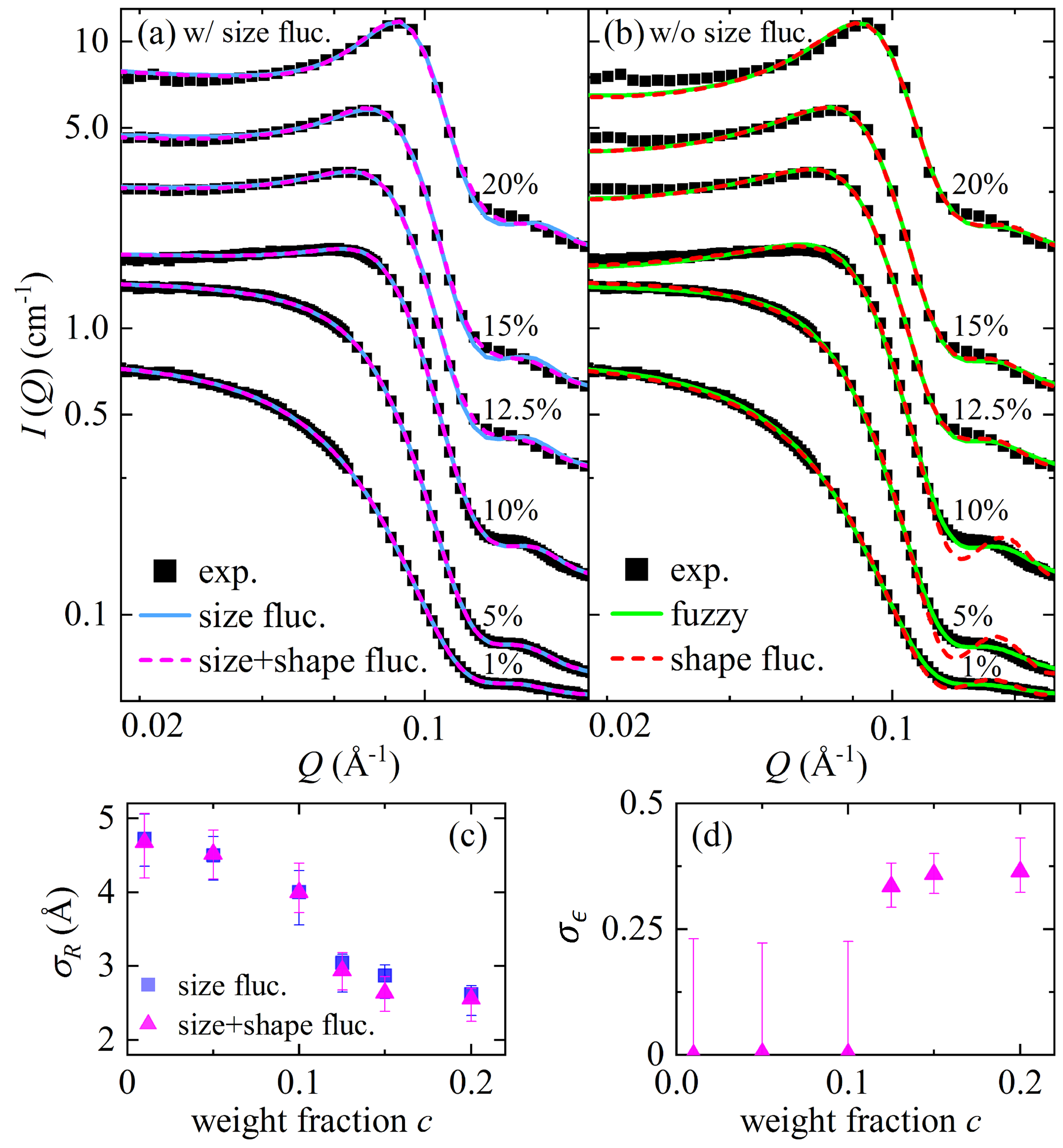}
	\caption{SANS analyses on the G6 PAMAM dendrimers dissolved in $\rm{D_2O}$ at $c = 1$, 5, 10, 12.5, 15 and 20 wt\%. Symbols in (a) and (b) denote the measured spectra(they are vertically shifted for visibility). (a) Fits with the size fluctuation model and the model with both the size and shape fluctuations. (b) Fits with the monodisperse fuzzy-ball model and the model only considering the shape fluctuation. These two models, without explicitly considering the size fluctuation,  underestimate the low-$Q$ intensity at $c\ge 10$ wt\%. (c) Size fluctuation ($\sigma_R$) of the G6 dendrimer as a function of $c$. The results are obtained with the size fluctuation model and the model with both the size and shape fluctuations. (d) Shape fluctuation ($\sigma_{\epsilon}$) of the G6 dendrimer as a function of $c$. The details of the SANS fitting can be found in the Supplementary Material.}
	\label{fig:Fig1}
\end{figure}

We first explore the size fluctuation by modeling the fluctuating dendrimers as a collection of polydisperse spheres. So, $P(Q)$ is expressed as \cite{Rathgeber2002, Bauer2001}:
\begin{eqnarray}
&& P_{\textrm{s}}(Q)=\sqrt{\frac{2}{\pi}} \frac{1}{\sigma_R} \left[ 1+\textrm{erf}\left(\frac{R}{\sqrt{2}\sigma_R}\right) \right]^{-1} \nonumber \\
&&\times
\int_0^{\infty} \left[ \frac{3 j_1(Qr)}{Qr} \right]^2 \textrm{exp}\left[-\frac{(r-R)^2}{2\sigma_R^2}\right] dr+a_{\textrm{b}}P_{\textrm{b}}(Q),\label{eq:Ps_2}
\end{eqnarray}
where $j_1(x)$ is the 1st order spherical Bessel function of the 1st kind, $P_{\rm b}(Q)$ represents the contribution from the intra-particle density variation with $a_{\rm b}$ denoting its amplitude \cite{Rathgeber2002, Daoud1982, Dozier1991}, $R$ is the average radius, and $\sigma_R$ is the standard deviation of the Gaussian distribution of $R$.  $\sigma_R$ reflects the fluctuating amplitude of the dendrimer size. This form factor ignores the “fuzzy” profile of the radial density distribution \cite{Rathgeber2002, Ballauff2004}. It could be reasonable for high-generation dendrimers, because their ratios of the fuzzy edge ($\sigma_{\rm f}$, see Eq. (\ref{eq:Pf_3})) to radius are relatively small \cite{Rathgeber2002}. The analyses for the samples of generation 6 (G6) are displayed in Fig. \ref{fig:Fig1} (a, blue lines), and are seen to be satisfactory at the measured concentrations.

To highlight the effect of the size fluctuation, we analyze the same data with the frequently-used monodisperse fuzzy-ball model \cite{Rathgeber2002, Pedersen2001,Chen2007}. Here, the size fluctuation is not considered, so $\beta(Q)=1$ and $S'(Q)=S(Q)$. $P(Q)$ is given by \cite{Rathgeber2002}:
\begin{equation}
P_{\textrm{f}}(Q)= \left[ \frac{3 j_1(QR)}{QR} \textrm{exp}\left(-\frac{Q^2\sigma_{\textrm{f}}^2}{4}\right)\right]^2 +a_{\textrm{b}}P_{\textrm{b}}(Q),\label{eq:Pf_3}
\end{equation}
where $\sigma_{\textrm{f}}$ denotes the spatial range of the fuzzy edge. This model has the same number of parameters as the preceding one. The fitting results, shown in Fig. \ref{fig:Fig1}(b), exhibit clear discrepancies from the measured spectra at $c\ge 10$ wt\%. The low-$Q$ intensity is remarkably underestimated. The peak position also deviates from the measured one. Comparing the two models, we assert that the size fluctuation is significant, and should be taken into account in analyzing the SAS data at $c\ge 10$ wt\%. Numerically, the incorporation of the size fluctuation lowers the low-$Q$ part of $\beta(Q)$ from 1 since $\beta(Q=0)=\langle r^3 \rangle^2/\langle r^6\rangle$ \cite{Chen1983}, and thus lifts the low-$Q$ part of $S'(Q)$ to match the experimental data \cite{beta1}. Notice that, at $c\le 5$ wt\%, both models work well, leading to contradictory results on the existence of the size fluctuation. This issue will be clarified  in the scattering contrast analysis shown later.

Considering that the dendrimer can deform to a spheroid-like form \cite{Goddard1989}, we incorporate the shape fluctuation by allowing the aspect ratio of a dendrimer ($\epsilon$) to fluctuate according to a Schultz distribution \cite{Chen1983} with the center at 1 and the standard deviation of  $\sigma_{\epsilon}$. The fitting results considering both the size and shape fluctuations are given in Fig. \ref{fig:Fig1}(a, magenta lines). It is seen that incorporating the shape fluctuation improves the fit at high $Q$. Notice that, solely incorporating the shape fluctuation cannot match the low-$Q$ spectra (Fig. \ref{fig:Fig1}(b)), since in this case $\beta(Q\to 0)$ does not deviate from 1 \cite{beta1}.

Figure \ref{fig:Fig1}(c) and (d) show the fitting results for $\sigma_R$ and $\sigma_{\epsilon}$ of the G6 dendrimer, respectively. At the dilute limit, $\sigma_R$ is 4.7 $\mathrm{\mathring{A}}$, corresponding to a dynamic polydispersity $\xi_R=\sigma_R/R$ of 15\%. As $c$ crosses about 11 wt\%, the size fluctuation is  suppressed while the shape fluctuation is strongly enhanced. In later part we will show that this behavior corresponds to a dynamic crossover.

As seen from Fig. \ref{fig:Fig1}, incorporating the size fluctuation is important  in decoding the structural information of ultrasoft colloids. Next, we will verify its existence from another view. Since dendrimer  has a water-accessible architecture \cite{Karatasos2001, Chen2012}, the contrast term $A$ in Eq. (\ref{eq:IQ_1}) is written as $A=\langle(b_{
\rm{pol}}+Nb_{\rm{w}}-n_{\rm{w}} V_{\rm{p}} b_{\rm{w}} )^2 \rangle$, where $b_{\rm{pol}}$ is the total scattering length of a dry dendrimer, $b_{\rm{w}}$ is the average scattering length of a water molecule, $N$ is the number of water molecules inside a dendrimer, $V_{\rm{p}}$ is the volume of a dendrimer in solution, $n_{\rm{w}}$ is the number density of bulk water, and $\langle \dots \rangle$ denotes the average over all particles. It is straightforward to find that:  
\begin{widetext}
\begin{equation}
A(b_{\rm w}) =  [ n_{\rm{w}}^2 (\langle V_{\rm p}\rangle^2+\langle \Delta V_{\rm p}^2\rangle) +\langle  N\rangle^2+\langle  \Delta N^2\rangle - 2 n_{\rm{w}} (\langle V_{\rm p}\rangle \langle N\rangle + \langle \Delta V_{\rm p} \Delta N\rangle)] b_{\rm w}^2- 2b_{\rm{pol}}(n_{\rm w}\langle V_{\rm p}\rangle -\langle N\rangle)b_{\rm w} + b_{\rm pol}^2,\label{eq:Abw_4}
\end{equation}
\end{widetext}
where $\Delta V_{\rm p}=V_{\rm p}-\langle V_{\rm p}\rangle$ and $\Delta N=N-\langle N\rangle$. $\Delta N$ and $\Delta V_{\rm p}$ should be highly correlated. To the first-order approximation, it can be assumed that $\Delta N$ is proportional to $\Delta V_{\rm p}$ by $\Delta N=n_{\rm{in}} \Delta V_{\rm p}$, where $n_{\rm{in}}$ should be smaller than $n_{\rm{w}}$ due to the excluded volume of the constituent atoms of dendrimer. So that Eq. (\ref{eq:Abw_4}) is rewritten as:
\begin{equation}
A(b_{\rm w})=(\alpha^2+\theta^2)b_{\rm w}^2 -2b_{\rm pol}\theta b_{\rm w}+b_{\rm pol}^2,\label{eq:Abw_5}
\end{equation}
where $\theta=n_{\rm w} \langle V_{\rm p}\rangle - \langle N\rangle$, and $\alpha=(n_{\rm w}/n_{\rm in} -1) \sigma_N$ with $\sigma_N=\sqrt{\langle\Delta N^2\rangle}$ denoting the fluctuation of the number of invasive water molecules. If no such fluctuation exists, then $\alpha=0$, and Eq. (\ref{eq:Abw_5}) reduces to:
\begin{equation}
A(b_{\rm w})=(\theta b_{\rm w} -b_{\rm pol})^2.\label{eq:Abw_6}
\end{equation}

\begin{figure}[h]
\centering
\includegraphics[scale=1]{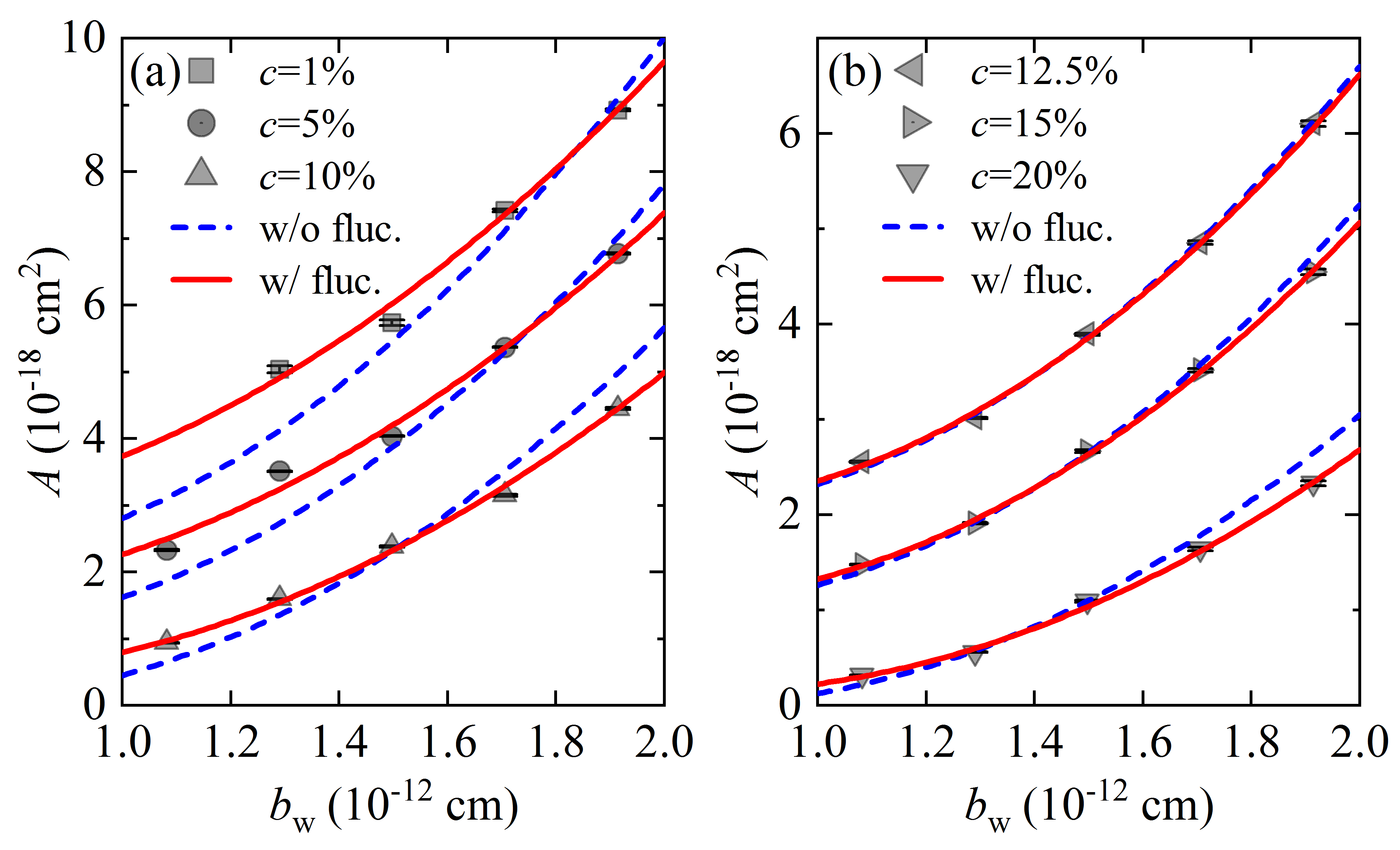}
\caption{Scattering contrast $A$ as a function of the average scattering length of solvent molecule $b_{\rm w}$ at $c\le 10$ wt\% (a) and $c> 10$ wt\% (b). Symbols denote the experimental results. Solid and dashed lines denote the fits using Eq. (\ref{eq:Abw_5}) (with size fluctuation) and Eq. (\ref{eq:Abw_6})   (without size fluctuation), respectively  \cite{contrast}. They are shifted vertically for visibility. }
\label{fig:Fig2}
\end{figure}

\begin{figure}[h]
\centering
\includegraphics[scale=1]{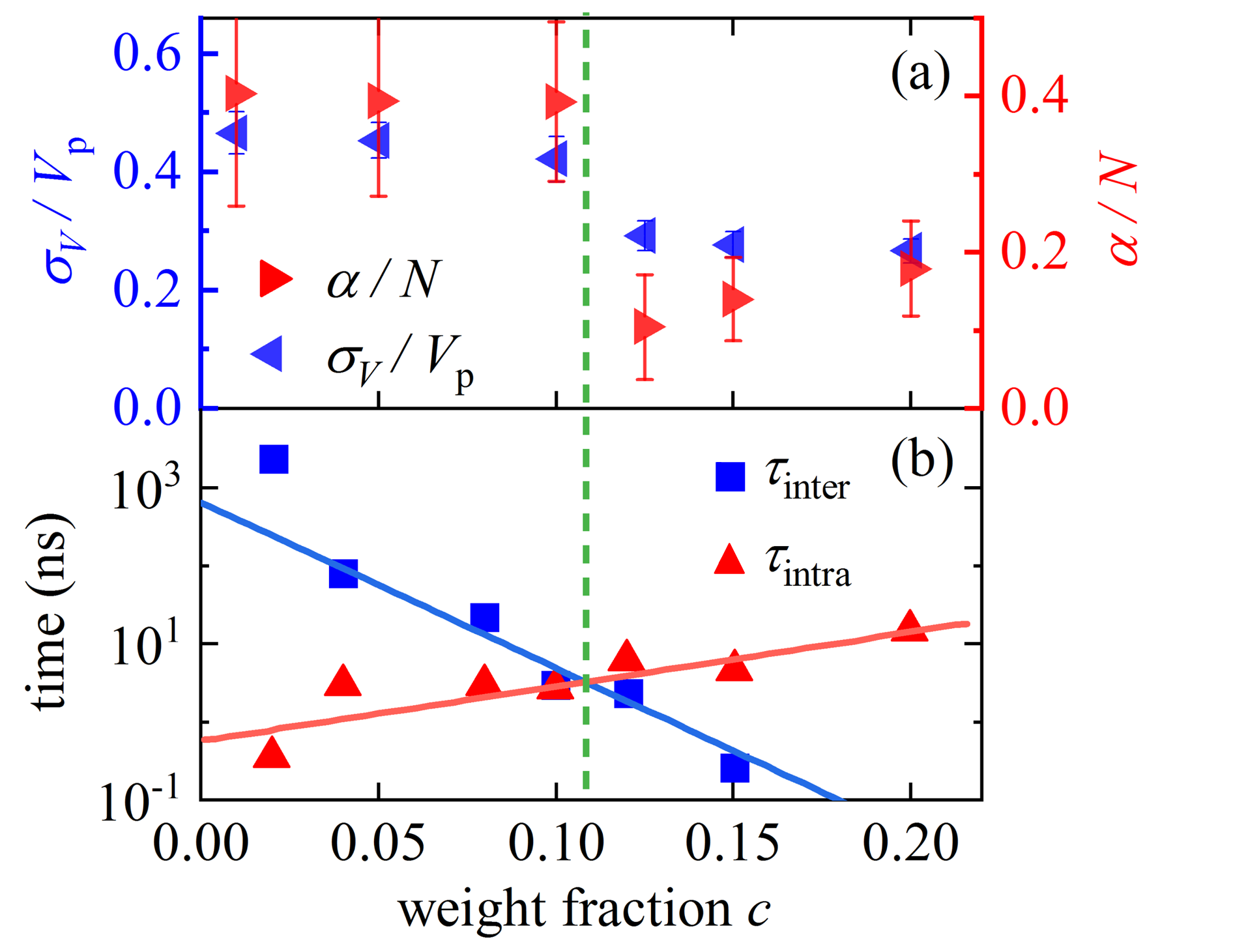}
\caption{(a) Fluctuations of volume ($\sigma_V/V_{\rm p}$) and invasive water ($\alpha/N \propto \sigma_N/N$) of a G6 dendrimer as a function of $c$. (b) Inter-particle collisional time $\uptau_{\rm{inter}}$ and intra-particle relaxation time $\uptau_{\rm{intra}}$ of G6 dendrimers as a function of $c$ \cite{Chen2014}. The dashed line marks the crossover concentration $c^*$.}
\label{fig:Fig3}
\end{figure}

Equations (\ref{eq:Abw_5}) and (\ref{eq:Abw_6}) provide an approach to verify the existence of the size fluctuation ($\propto \sigma_N$) from a microscopic view. One can vary $b_{\rm w}$ by changing the molar fraction of $\rm{D_2O}$ in solvent ($\gamma$), and fit the experimental $A(b_{\rm w})$ with Eqs.(\ref{eq:Abw_5}) and (\ref{eq:Abw_6}). If the size fluctuation is considerable, the fitting quality with Eq. (\ref{eq:Abw_5}) will be better than that with Eq. (\ref{eq:Abw_6}). We vary $\gamma$ from 100\% to 60\% for all concentrations, and fit the experimental contrast term with Eqs.(\ref{eq:Abw_5}) and (\ref{eq:Abw_6}). The results are shown in Fig. \ref{fig:Fig2}. It is seen that at $c\le 10$ wt\%, Eq. (\ref{eq:Abw_5}) performs much better than Eq. (\ref{eq:Abw_6}), suggesting that the fluctuations of size and invasive water exist and strongly affect the scattering contrast. At $c>10$ wt\%, this effect is less significant, implying a smaller fluctuation. The values of $\alpha/N$, representing the fluctuation of invasive water, are given in Fig. \ref{fig:Fig3}(a). In the same panel, we also plot the particle volume fluctuation $\sigma_V/V_{\rm p}$ obtained from the model fitting of SANS spectra as illustrated in Fig. \ref{fig:Fig1}(a). It is remarkable that these two quantities display very similar behaviors, especially considering that they are found with different approaches \cite{contrast}. Such consistency confirms the existence of the size fluctuation in the G6 dendrimer.

As mentioned above, SAS analyses on neutral ultrasoft colloids usually deteriorate when $c$ is larger than about 10 wt\%. It has been tentatively attributed to the failure of factorizing $I(Q)$ into the product of $P(Q)$ and $S(Q)$ due to the inter-particle interpenetration or overlap \cite{Likos2005, Likos1998, Ballauff2004}. Nevertheless, many studies show that strong interpenetration or overlap does not occur at such low concentrations \cite{Topp1999, Jackson1998, Rietveld2000, Pedersen2015, Zacharopoulos2002, Uppuluri1998, Schu2017}. Our analysis suggests that the failure of $P(Q)\cdot S(Q)$ factorization is due to the molecular fluctuation, and can be corrected with Eq. (\ref{eq:IQ_1}). In fact, Pedersen also found that the SANS intensity of block copolymer micelles is expressed by Eq. (\ref{eq:IQ_1}) rather than a product of $P(Q)$ and $S(Q)$ by considering the configuration distribution of the chains in corona \cite{Pedersen2001}. We argue that the particle conformational fluctuation is the basis for both Ref. \cite{Pedersen2001} and our work. This agreement suggests that Eq. (\ref{eq:IQ_1}) is the better expression to account for the softness, which originates from the flexible architecture, of ultrasoft particles such as dendrimers and starlike polymers. Note that, Eq. (\ref{eq:IQ_1}) is based on the assumption  that the particle conformation decouples with the position \cite{Chen1983}. It becomes invalid at concentrations close to random close packing because of the considerable spatial heterogeneity of particle deformation \cite{Zaccarelli2019}.

As seen from Fig. \ref{fig:Fig1} and Fig. \ref{fig:Fig3}(a), all discussed fluctuations exhibit crossovers at $c^*\approx 11$ wt\%. Here, we seek for the dynamic origin of this phenomenon. Figure \ref{fig:Fig3}(b) shows the results of an NSE study on the same system \cite{Chen2014}. By decomposing the motion of a dendrimer into the translational diffusion and the internal relaxation, this NSE analysis gives two characteristic times: the inter-particle collisional time $\uptau_{\rm inter}$ and the intra-particle relaxation time $\uptau_{\rm intra}$. Interestingly, $\uptau_{\rm inter}$ and $\uptau_{\rm intra}$ also intersect at $c^*$. So, we associate the fluctuation crossover with the dynamic process. At $c<c^*$, conformational fluctuations emerge due to the enormous internal degrees of freedom. The inter-particle collisions do not strongly perturb the internal relaxation since $\uptau_{\rm intra}<\uptau_{\rm inter}$. While at $c>c^*$, $\uptau_{\rm intra}>\uptau_{\rm inter}$, the frequent collisions hinder the dendrimer from fully exploring its conformational space at the dilute limit, and thus restrain the size fluctuation. Meanwhile, these collisions enhance the particle deformation and suppress the internal relaxation, which results in an increasing shape fluctuation.

\begin{figure}[h]
\centering
\includegraphics[scale=1]{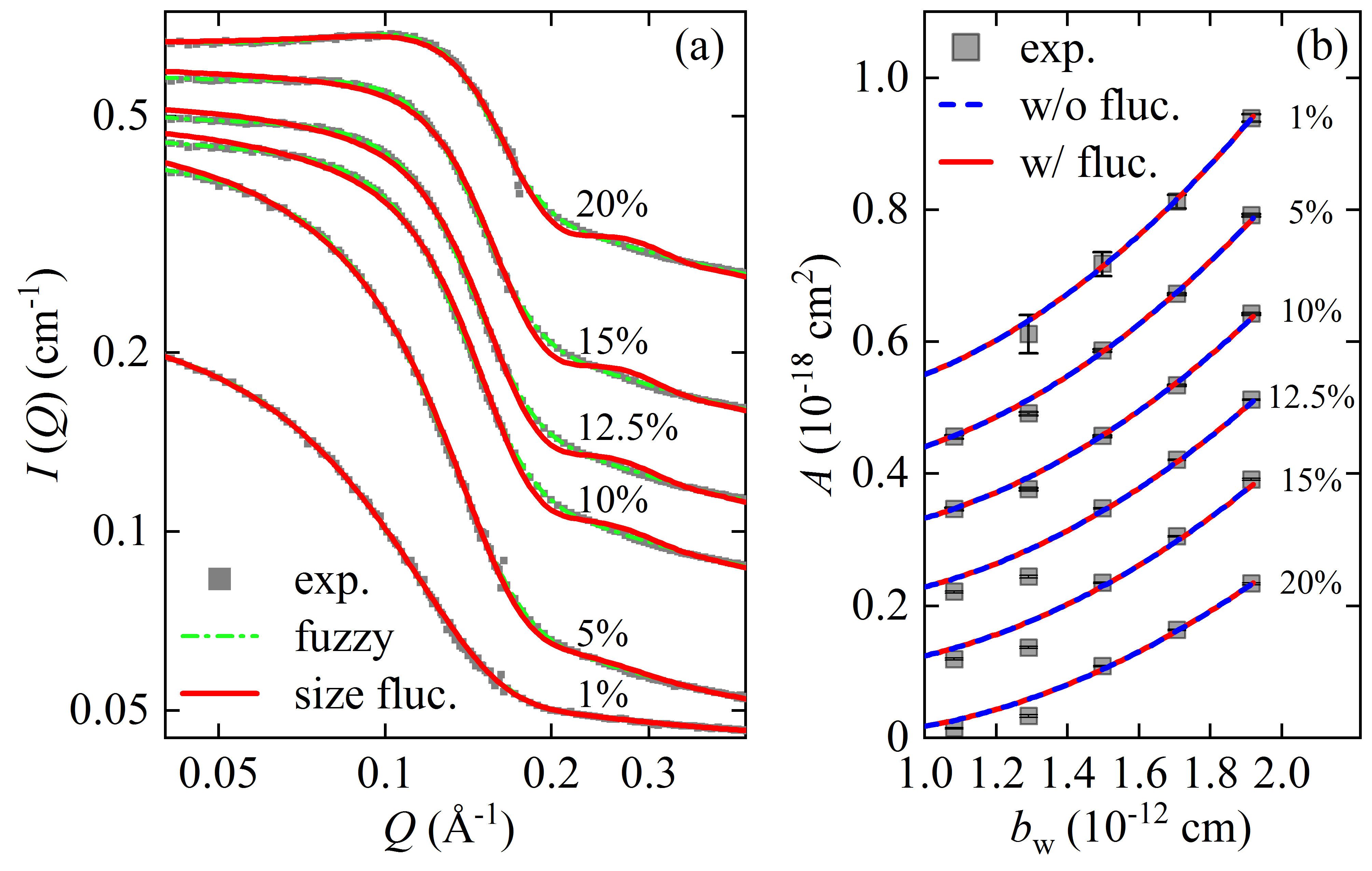}
\caption{(a) SANS analysis of G4 PAMAM dendrimers dissolved in $\rm D_2O$ at $c = 1$, 5, 10, 12.5, 15 and 20 wt\%. Symbols denote the measured spectra. Solid and dashed lines respectively denote the fits using the monodisperse  fuzzy-ball model and the size fluctuation model. (b) Scattering contrast of G4 dendrimers as a function of $b_{\rm{w}}$. Symbols denote the experimental results. Solid and dashed lines respectively denote the fits using Eq. (\ref{eq:Abw_5}) and Eq. (\ref{eq:Abw_6}). Data are vertically shifted.}
\label{fig:Fig4}
\end{figure}

We further measure the samples of generation 4 (G4) to investigate the structural origin of the observed fluctuations. The polydisperse sphere model and monodisperse fuzzy-ball model are applied to fit the SANS spectra and the results are shown in Fig. \ref{fig:Fig4}(a). In contrast to the G6 samples, for G4 samples the fuzzy-ball model works well, while the polydisperse sphere model overestimates the low-$Q$ intensity, and gives an inappropriate oscillation at high $Q$. This result indicates that the size fluctuation is imperceptible for G4 samples. To confirm this observation, we perform the contrast analysis given by Eqs.(\ref{eq:Abw_4})-(\ref{eq:Abw_6}). As seen from Fig. \ref{fig:Fig4}(b), Eqs.(\ref{eq:Abw_5}) and (\ref{eq:Abw_6}) give the same fits, showing that the size fluctuation term $\alpha$ equals to zero. The sharp difference between the G4 and G6 samples reveals the role of molecular structure in determining the fluctuations. For our G4 sample, the ratio of the fuzzy edge to radius ($\sigma_{\textrm{f}}/R$) is 74\%, which demonstrates the pronouncing open feature of its periphery. Intuitively, the strong fuzzyness implies an unsharp boundary, which makes the particle volume and its fluctuation not so well-defined. On the contrary, G6 dendrimers are found to be alike spheres with a clearer boundary. This result agrees with the previous finding that the dendrimer structure evolves from an open one to a compact one as the generation increases \cite{Goddard1989, Karatasos2001, Rathgeber2002, Paulo2007, Goddard2004}. The evident compactness in G6 dendrimers leads to small intra-particle free volume, which enhances the interaction between dendrons \cite{Mansfield1994, Zacharopoulos2002, Ballauff2004}. Therefore, it is reasonable that the dendrons can fluctuate collectively with long-range correlation, and form the size fluctuation of the whole particle. For G4 dendrimers, a simulation shows that the internal fluctuations are uncorrelated at large distances \cite{Likos2003} due to the open structure. In this case, intra-particle motions involving long-range correlated motion of monomers, such as the size fluctuation, should be weak. In addition, an NSE study suggests that the molecular fluctuation of low-generation starlike dendrimers is dominated by the breathing mode that does not induce significant size and shape fluctuations \cite{Rathgeber2006}. This feature is also attributed to the fuzzy nature of the radial density profile of the molecule \cite{Rathgeber2006}.

In summary, we investigate the fluctuation of ultrasoft colloids by using PAMAM dendrimers as the model system. By considering the dynamic polydispersity and the fluctuation of invasive water in SANS analysis, we reveal the fluctuating amplitudes of the size and shape of the G6 dendrimer. The size fluctuation is found to be of particular importance in interpreting the SAS data. These fluctuations exhibit strong dependences on the dynamics and structure. The competition between the intra- and inter-particle dynamics introduces a crossover concentration above which the size fluctuation is suppressed, and the shape fluctuation increases. The absence of size fluctuation in G4 dendrimers highlights the importance of the structural compactness in determining the intra-particle motion. Our approach provides a basis for further exploration on the fluctuation effect on various structural and dynamic properties of ultrasoft colloids.


\begin{acknowledgments}
This research was supported by the National Natural Science Foundation of China (No. 11975136, U1830205). We thank the EQ-SANS at Spallation Neutron Source, Oak Ridge National Lab and the D22 beamline at Institut Laue-Langevin for beamtime.
\end{acknowledgments}

\bibliography{refs}

\begin{thebibliography}{60}
\bibitem [{bet()}]{beta1}%
  \BibitemOpen
  \href@noop {} {\bibinfo {title} {It can be seen from {F}ig. {S}4 in the
  {S}upplementary {M}aterial.}}\BibitemShut {Stop}%
\end{thebibliography}

\end{document}